# Outburst characteristics of the dwarf nova V452 Cassiopeiae

Jeremy Shears, Christopher Lloyd, David Boyd, Steve Brady, Ian Miller and Roger Pickard

**Abstract**

V452 Cas was thought to have rare outbursts, but monitoring from 2005 to 2008 has shown that the outburst interval is about one month and is weakly periodic. Observations of seven superoutbursts over the same period shows a very repeatable superoutburst period of 146±16 days. Time series photometry of the 2007 September superoutburst shows that the outburst reached magnitude 15.3 at maximum and had an amplitude of 3.2 magnitudes. The outburst lasted for 12 days. Early superhumps with an amplitude of 0.3 magnitudes and period of $P_{sh}$ = 0.08943(7) days gave way to superhumps with decreasing amplitude and $P_{sh}$ = 0.08870(2) days later in the outburst, corresponding to a continuous period change $\dot{P}/P = -9(2) \times 10^{-4}$ d$^{-1}$. V452 Cas has one of the smallest outburst amplitudes and shortest superoutburst periods of typical UGSU systems.

**Introduction**

Dwarf novae are a class of cataclysmic variable star in which a white dwarf primary accretes material from a secondary star via Roche lobe overflow. The secondary is usually a late-type main-sequence star. In the absence of a significant white dwarf magnetic field, material from the secondary is processed through an accretion disc before settling on the surface of the white dwarf. As material builds up in the disc, a thermal instability is triggered that drives the disc into a hotter, brighter state causing an outburst in which the star brightens by several magnitudes [1]. Dwarf novae of the SU UMa family (UGSU) occasionally exhibit superoutbursts which last several times longer than normal outbursts and may be up to a magnitude brighter. During a superoutburst the light curve of a UGSU star is characterised by superhumps. These are modulations in the light curve which are a few percent longer than the orbital period. They are thought to arise from the interaction of the secondary star orbit with a slowly precessing eccentric accretion disc. The eccentricity of the disc arises because a 3:1 resonance occurs between the secondary star orbit and the motion of matter in the outer accretion disc. For a more detailed review of UGSU dwarf novae and superhumps, the reader is directed to references 1 and 2.

**History of V452 Cas**

V452 Cas was first identified as a dwarf nova by Richter on plates taken at the Sonneberg observatory in 1967 [3]. He found it "bright" between Nov 3 and 8, whereas on Oct 28 it was fainter; he lists the range as 14p to ~17.5p. Bruch *et al.* photographed the field of V452 Cas on 13 nights between 1980 and 1983 and found it once in outburst [4]. Liu and Hu performed spectroscopy in 1999 Jan when the star was in quiescence at 18.6V, which confirmed it to be typical of a dwarf nova [5]. The AAVSO International Database lists 1563 individual observations of the star between the first entry on 1989 Aug 28 and 2005 Feb 20, but only 29 of these are positive detections and they probably relate to eight separate outbursts as given in Table 1. The maximum reported magnitude for these outbursts is between 14.7 and 15.8, but since the majority of observations in the database have a limiting magnitude of ~14.5, it is likely that many outbursts have been missed as the star is fainter than this in outburst. As rather few outbursts of V452 Cas had been seen, it was generally assumed to be an infrequently outbursting dwarf nova and it was added to the BAA Variable Star Section's Recurrent Objects Programme [6]. The aim of this programme is to encourage observations of poorly studied eruptive stars where outbursts occur at periods of greater than a year.

V452 Cas was identified as a member of the UGSU family of dwarf novae as a result of photometry carried out by Tonny Vanmunster and Robert Fried during the 1999 Nov outburst which showed characteristic superhumps with a period, $P_{sh}$ = 0.0891(4) d. Several references quote a shorter period (e.g. ref 7,8), but this seems to have been based on preliminary reports by the two observers [9]. Superhumps were also reported during the 2000 Sep outburst confirming this as a superoutburst [10].

**Outburst period**

In an effort to characterise the outburst behaviour of V452 Cas the authors undertook regular monitoring between 2005 Aug 15 and 2008 Apr 9, via CCD imaging. The advantage of this programme is a more consistent coverage and a much fainter limiting magnitude, typically 17.2C, than was achieved previously. Where appropriate, these observations have been supplemented by additional data from the AAVSO database. During the three seasons covered, a total of 23 outbursts have been detected and the details of these are given in Table 2.

It is clear from the outburst durations listed in Table 2 that there are two categories of outburst exhibited by V452 Cas. Sixteen of the outbursts were short, lasting less than 4 days and were never brighter than 16.0C. The remaining seven were longer, lasting more than 8 days and reached 15.8C or brighter. According to AAVSO observations, the 1999 confirmed superoutburst lasted 12 days, therefore we interpret the category of short outbursts as "normal" outbursts and the longer outbursts as superoutbursts. This categorisation is further supported by our confirmation that the long outburst in 2007 Sep was also a superoutburst, to be discussed later, as was the poorly observed superoutburst in 2008 Feb [11]

The intervals, $\Delta T$, between consecutively detected outbursts are given in Table 2 and the distribution is plotted as a histogram in Figure 1. Gaps in the data mean that some outbursts will be missed, but given the generally good coverage this plot should give an indication of the likely outburst time scale. The distribution of outburst intervals appears to be bi-modal with the highest concentration around 27 days and another cluster around 33 days, as well as a wide distribution of other intervals. The longest intervals are due to the seasonal gap in the data and those above 40 days may represent multiples of a few outbursts cycles. The median interval is 34 days. The shortest spacing of 17 days is seen only once and is generated by a single faint sighting. It seems unlikely that the outburst interval is any shorter than this as the median of data spacing is less than 2 days so intervals under 25 days are well sampled.

Although the plot of the intervals is instructive it does not indicate a completely dominant time scale nor does it help to identify any periodic behaviour in the outbursts. To identify potential periods the Discrete Fourier Transform (DFT) power spectrum of the positive observations has been constructed and is shown in Figure 2a. The main features lie between f=0.02 and 0.04 c d$^{-1}$, 25 to 50 days, which corresponds to the dominant range of outburst intervals. There is also another feature near f=0.007 c d$^{-1}$ or 143 days, which will become clear later. To approach this from another direction the "Window function" of the times of the outburst detections (as opposed to all the data) has been calculated and is shown in Figure 2b. The window function is normally used in combination with the DFT power spectrum to help identify spurious periods that are due to the data spacing alone. However, when used on the times of the outbursts it should reveal any periodicities in those timings. The only real feature lies at f=0.033 c d$^{-1}$ or 31 days, which is between the two main clusters in Figure 1 and Figure 2a. It is relatively weak, suggesting a poor periodicity, which is what might be expected given what is known about the outburst intervals. Using this as an initial period a linear ephemeris has been calculated for all the outbursts giving,

$$JD_{Max} = 2453642(2) + 30.7(2) \times E \ldots\ldots(1)$$

The window function has also been calculated for just the superoutburst timings and this is shown in Figure 2c. The most prominent peak lies at f=0.007 c d$^{-1}$ or 143 days, and corresponds to the same feature seen in the power spectrum in Figure 2a. The linear ephemeris for the superoutbursts is,

$$JD_{Max} = 2453625(12) + 146(4) \times E \ldots\ldots(2)$$

The corresponding O-C diagrams are shown in Figure 3a and 3b and it is clear that while there is some coherence in the outburst timings there is also considerable scatter in the individual outburst intervals. The O-C residuals range over ±10 days, which is one third of the outburst period. It is also clear that the superoutbursts do not behave differently from normal outbursts in Figure 3a. There is rather larger scatter in the superoutburst timings, but as the period is much longer, the fractional change is smaller, giving the superoutbursts a much more regular clock. Although the formal error on the superoutburst period is small, a more realistic idea of when one might be seen is given by the standard deviation of the O-C residuals, so the adopted superoutburst period is 146±16 days.

The ratio of the superoutburst to normal outburst period 4.8±0.3, so is consistent with 5, but superoutbursts do not occur every fifth outburst. In the normal outburst cycles the superoutbursts occur at cycles 0, 4, 9, 13, 18, 23 and 29, giving intervals of 4, 5, 4, 5, 5 and 6 outbursts. Unfortunately, except in the first and last cases, it is not possible to simply count the number of normal outbursts per superoutburst as some faint outbursts have still probably been missed despite the intensive monitoring.

The uncertainty on the superoutburst period is sufficiently large that the ephemeris cannot reliably be projected back to the period covered by the earlier outbursts given in Table 1. However, the confirmed superoutbursts in 1999 Nov [9] and 2000 Sep [10] can be used to improve the ephemeris and identify other potential superoutbursts from around that time. These appear to be the 1999 Jun outburst, which appears 144 days earlier and probably the 1997 Nov outburst, and the 1997 Aug outburst, but unfortunately these have only one observation each. The extended superoutburst ephemeris is

$$JD_{Max} = 2453637(6) + 142(1) \times E \ldots\ldots(3)$$

and the standard deviation of the residuals is 17 days. In this ephemeris and the previous one given in equation 2 the residuals are dominated by the timing of the most recent superoutburst which seems to have occurred very late. The only other estimate of the superoutburst period is 320/n days suggested by Kato *et al.*[7], which comes from the interval between the 1999 Nov [9] and 2000 Sep [10] superoutbursts, with n=2..

**The 2007 September superoutburst**

The outburst of V452 Cas discussed in this paper was detected on Sep 1.843 at 15.7C by JS, the previous observation have been made 4 days earlier on Aug 28.226 when the star was undetectable (<17.2C). Twenty unfiltered time-series photometry runs were conducted during the course of the outburst yielding more than 127 hours of data. Table 3 summarises the instrumentation used and Table 4 contains a log of the time-series runs. In all cases raw images were flat-fielded and dark-subtracted, before being analysed using commercially available differential aperture photometry software against the comparison star sequence given in the AAVSO chart 020301 [12]. From a comparison of overlapping runs, any systematic differences between observers are similar to the errors. An image of the star in outburst is shown in Figure 4. We shall frequently refer to dates in the truncated form JD = JD – 2454000.

The overall light curve of the outburst is shown in the bottom panel of Figure 5. The first 4 days appears to correspond to the plateau phase commonly seen near the beginning of dwarf nova outbursts, although there was a very gradual brightening which suggest that the outburst had not long been underway when it was detected. At its brightest it was mag 15.3. This was followed by a small dip in the light curve and a subsequent partial recovery. The star then showed an approximately linear decline at 0.16 mag/d from JD 351 to 356. Finally there was a sharp decline to quiescence at about magnitude 18.5, twelve days after the outburst was detected. Thus the outburst amplitude was a very modest 3.2 magnitudes. Although it was much less well observed the 2008 Feb superoutburst followed a very similar pattern.

Some of the longer photometry runs are shown in Figures 6a to g. Superhumps having peak-to-peak amplitude of ~0.3 mag were obvious on the first night of discovery, confirming this to be a superoutburst. The amplitude of the superhumps remained constant for the first 6 days of the outburst, covering the period from discovery up to the initial phase of the decline (JD 345 to 351). As the decline continued, the superhumps diminished in amplitude to ~0.2 mag on JD 354 and ~0.1 mag on the final night of observation, although by this time they were hard to distinguish above the noise.

**Measurement of the superhump period**

To study the superhump behaviour, we first extracted the times of each sufficiently well-defined superhump maximum from the individual light curves according to the Kwee and van Woerden method [13] using the Peranso software [14], although we recognise that the errors based on this method appear to be underestimates. Times of 24 superhump maxima were found (Table 5). Following a preliminary assignment of superhump cycle numbers to these maxima, an analysis of the times of maximum indicated that the superhump period appeared to remain constant for the first 21 cycles, after which it changed. We obtained the following linear superhump maximum ephemeris for the interval JD 345 to 347:

$$JD_{Max} = 2454345.5992(11) + 0.08943(7) \times E$$

The observed minus calculated (O–C) residuals relative to this ephemeris are plotted in the top panel of Figure 5. This shows that sometime between JD 347 and 348 (cycle 21 and 31) the superhump period changed. Analysing the times of maximum from JD 348 to 354 (cycles 31 to 102) yielded a linear superhump maximum ephemeris:

$$JD_{Max} = 2454345.6132(20) + 0.08870(2) \times E$$

Thus in the early part of the outburst the superhump period, $P_{sh}$, was 0.08943(7) d, subsequently decreasing to $P_{sh}$ = 0.08870(2) d. A reduction in $P_{sh}$ has been observed in many UGSU stars as the outburst progresses, including DV UMa and IY UMa [15,16] and may be explained by the accretion disc emptying and shrinking. Whether in the case of V452 Cas the period changed abruptly or more gradually is difficult to say due to the lack of data during the critical period. However, the data are also consistent with a continuous change in period with $\dot{P}/P = -9(2) \times 10^{-4}$ $d^{-1}$, but the residuals are about 1.5 times larger than with the abrupt change. The period change seems to correspond to the end of the plateau period of the outburst and there is a suggestion of a concomitant change in the superhump profiles to a less symmetrical shape, with a slower decline than rise.

To confirm our measurement of $P_{sh}$, we carried out a period analysis of the combined data from JD 345 to 347 using the Data Compensated Discrete Fourier Transform (DCDFT) algorithm in Peranso, after subtracting the mean and linear trend from each of the light curves. This gave the power spectrum in Figure 7 which has its highest peak at a period of 0.0894(4) d (other peaks are 1d aliases), which we interpret as $P_{sh}$; this value is consistent with our earlier measurement. The superhump period error estimate is derived using the Schwarzenberg-Czerny method [17]. Several other statistical algorithms in Peranso gave the same value of $P_{sh}$. In a similar manner we analysed the data from JD 348 to 354, which yielded $P_{sh}$ = 0.0886(3) d (and its 1 day aliases; power spectrum not shown), which is again consistent with the result from analysing times of superhump maxima.

Removing $P_{sh}$ from each of the two power spectra leaves only weak signals, none of which has any significant relationship to the superhump or orbital periods. We searched for signals that could be associated with an orbital hump, and which could therefore yield the hitherto unknown orbital period, but none were forthcoming. Phase diagrams of the data from each section of the superhump regime, folded on their respective values of $P_{sh}$, are shown in Figure 8 and 9, where two cycles are shown for clarity. This suggests that $P_{sh}$ remained constant during each of the two regimes. We note that both values of the superhump period are consistent with the value of 0.0891(4) d reported by Vanmunster during the 1999 Nov superoutburst of V452 Cas [9].

**Position of V452 Cas**

In the past there has been some confusion about the position of V452 Cas, initially because an incorrect star was identified during astrometry at quiescence. Another factor may have been the presence of a close field star of mag 15.8 .The problem was resolved by Liu and Hu who identified the star in quiescence via spectroscopy [5]. We measured the position of V452 Cas on each of 6 images from Sep 3 when it was at its brightest in the superhump cycle using the Astrometrica astrometry software [18] and the USNO-B1.0 catalogue. It was well resolved from the nearby star on all these images. All 6 images gave exactly the same position in RA and with a variance of only 0.1" in Dec. The mean position is:

00h 52m 17.98s +53d 51' 50.1"   (J2000)

Errors in both RA and Dec from Astrometrica/USNO-B1.0 are +/-0.2". Our position is therefore almost identical to that given in Downes *et al.* [19] of 00h 52m 18.06s +53d 51' 50.1" (J2000).

**Discussion**

The outburst amplitude during the 2007 Sep superoutburst was only 3.2 magnitudes which is at the lower end of the distribution of outburst amplitudes for the vast majority of UGSU stars. The superoutburst period at 146 days is also at the lower end of the usual range and the only stars that show lower amplitudes and shorter superoutburst periods are the ER UMa subclass of UGSU variables. The search for "transition" objects, if such things exist, or perhaps more properly "intermediate" objects between typical UGSU systems and ER UMa stars has so far drawn a blank and unfortunately V452 Cas is not a contender. The superhump periods, and therefore the orbital periods of ER UMa systems, are generally < 0.07 d, so are among the shortest seen for UGSU stars and overlap those of the WZ Sge subclass. For V452 Cas $P_{sh}$ = 0.089 d, which is longer than the majority of UGSU systems, but it is still typical of the class. We note that YZ Cnc has some similar characteristics to V452 Cas, with $P_{sh}$ = 0.092 d and a superoutburst period of 134 days [7].

**Conclusions**

A monitoring programme of V452 Cas covering the past three observing seasons has shown that, in contrast to earlier belief, the outburst interval is relatively short at about one month and there is evidence that this is weakly periodic. Superoutbursts observed over the same period have a well-defined outburst period of 146 ± 16 days.

Time resolved photometry during the 2007 Sep superoutburst shows that the first four days corresponded to the plateau phase, with very gradual brightening to magnitude 15.3 at maximum. This was followed by a small dip in the light curve and a subsequent recovery, leading to an approximately linear decline at 0.16 mag/d for the last six days. Finally there was a sharp decline to quiescence at about magnitude 18.5, twelve days after the outburst was detected. The outburst amplitude was 3.2 magnitudes. Superhumps with a peak-to-peak amplitude of ~0.3 mag were visible for the first 6 days of the outburst, covering the period from discovery up to the initial phase of the decline. As the decline continued, the superhump amplitude also diminished to about ~0.1 mag on the final night of observation.

During the early part of the outburst the superhump period $P_{sh}$ = 0.08943(7) d, subsequently decreasing to $P_{sh}$ = 0.08870(2) d. The period change seems to correspond to the end of the plateau period of the outburst and a possible change in the superhump profiles. The data are also consistent with a continuous change in period with $\dot{P}/P = -9(2) \times 10^{-4}$ $d^{-1}$.

V452 Cas has one of the smallest outburst amplitudes and shortest superoutburst periods of typical UGSU systems.

**Acknowledgements**

The authors gratefully acknowledge the use of observations from the AAVSO International Database contributed by observers worldwide, the use of SIMBAD, operated through the Centre de Données Astronomiques (Strasbourg, France), and the NASA/Smithsonian Astrophysics Data System. We thank Tonny Vanmunster for providing information on the 1999 superoutburst of V452 Cas conducted by himself and Robert Fried. Finally, we are indebted to the referee whose suggestions have improved the paper.

**Addresses:**
JS: "Pemberton", School Lane, Bunbury, Tarporley, Cheshire, CW6 9NR, UK [bunburyobservatory@hotmail.com]
CL: Open University, Milton Keynes, MK7 6AA, UK [C.Lloyd@open.ac.uk]
DB: 5 Silver Lane, West Challow, Wantage, Oxon, OX12 9TX, UK [drsboyd@dsl.pipex.com]
SB: 5 Melba Drive, Hudson, NH 03051, USA [sbrady10@verizon.net]
IM: Furzehill House, Ilston, Swansea, SA2 7LE, UK [furzehillobservatory@hotmail.com]
RP: 3 The Birches, Shobdon, Leominster, Herefordshire, HR6 9NG [rdp@astronomy.freeserve.co.uk]

| Detection date | JD | Magnitude at maximum | Super-outburst |
|---|---|---|---|
| 1993 Oct 10 | 2449271.3 | 15.2Vis | |
| 1997 Aug 16 | 2450676.5 | 15.8C | Y? |
| 1997 Nov 18 | 2450770.6 | 14.58R | Y? |
| 1999 Jun 19 | 2451348.5 | 15.5Vis | Y? |
| 1999 Oct 5 | 2451456.8 | 14.7Vis | |
| 1999 Nov 9 | 2451492.3 | 14.7Vis | Y |
| 2000 Sep 20 | 2451808.4 | 15.1Vis | Y |
| 2002 Nov 2 | 2452581.5 | 15.3Vis | |

**Table 1: Outbursts of V452 Cas prior to 2005**
Data are taken from the AAVSO International Database C= unfiltered CCD, R = CCD + Johnson R filter, Vis = visual. The confirmed superoutbursts are based on the detection of superhumps from time series photometry during the 1999 Nov [9] and 2000 Sep [10] outbursts.

| Detection date (UT) | JD (-2450000) | ΔT (d) | Magnitude at maximum (C) | Outburst duration (d) | Super-outburst |
|---|---|---|---|---|---|
| 2005 Sep 23 | 3637.4 | | 15.7 | >15 | Y |
| 2005 Nov 3 | 3677.7 | 40.3 | 16.6 | <2 | |
| 2005 Nov 28 | 3703.3 | 25.6 | 16.2 | <3 | |
| 2005 Dec 26 | 3731.3 | 28.0 | 16.0 | <3 | |
| 2006 Jan 29 | 3765.3 | 34.0 | 15.8 | >8 | Y |
| 2006 Mar 18 | 3813.3 | 48.0 | 16.3 | <4 | |
| 2006 Apr 4 | 3830.4 | 17.1 | 17.2 | <4 | |
| 2006 Jul 12 | 3928.5 | 98.1 | 15.7 | >14 | Y |
| 2006 Aug 6 | 3953.6 | 25.1 | 16.5 | <3 | |
| 2006 Aug 28 | 3976.4 | 22.8 | 16.6 | <3 | |
| 2006 Oct 2 | 4011.3 | 34.9 | 16.4 | <5 | |
| 2006 Nov 5 | 4045.3 | 34.0 | 15.8 | >12 | Y |
| 2007 Jan 5 | 4106.3 | 61.0 | 16.0 | <4 | |
| 2007 Feb 27 | 4159.3 | 53.0 | 16.0 | <3 | |
| 2007 Apr 1 | 4192.3 | 33.0 | 16.0 | >8 | Y |
| 2007 Jul 30 | 4312.4 | 120.1 | 16.0 | <3 | |
| 2007 Sep1 | 4344.9 | 32.5 | 15.3 | >11 | Y |
| 2007 Oct 9 | 4383.3 | 38.4 | 16.6 | <4 | |
| 2007 Nov 13 | 4417.6 | 34.3 | 16.4 | <3 | |
| 2007 Dec 10 | 4445.4 | 27.8 | 16.9 | <4 | |
| 2008 Jan 5 | 4471.2 | 25.8 | 16.4 | <3 | |
| 2008 Feb 1 | 4498.4 | 27.2 | 16.2 | <4 | |
| 2008 Feb 28 | 4525.4 | 27.0 | 15.8 | >10 | Y |

**Table 2: Outbursts of V452 Cas between 2005 Aug and 2008 Apr**
Data from the authors and the AAVSO International Database. The superoutbursts are identified as being brighter than 16.0C and lasting longer than 8 days, while the 2007 Sep superoutburst also has confirmed superhumps.

| Observer | Telescope | CCD (unfiltered) |
|----------|-----------|------------------|
| JS | 0.28 m SCT | Starlight Xpress SXV-M7 |
| DB | 0.35 m SCT | Starlight Xpress SXV-H9 |
| SB | 0.4 m reflector | SBIG ST-8XME |
| IM | 0.35 m SCT | Starlight Xpress SXVF-H16 |
| RP | 0.30 m SCT | Starlight Xpress SXV-H9 |

**Table 3: Equipment used**

| Date in 2007 (UT) | Start time (JD-2454000) | Duration (h) | Observer |
|-------------------|-------------------------|--------------|----------|
| Sep 2  | 345.564 | 7.3 | SB |
| Sep 3  | 346.566 | 7.4 | SB |
| Sep 3  | 347.345 | 4.0 | JS |
| Sep 3  | 347.364 | 4.7 | RP |
| Sep 3  | 347.383 | 3.3 | DB |
| Sep 4  | 348.343 | 4.2 | IM |
| Sep 4  | 348.354 | 2.4 | RP |
| Sep 6  | 350.330 | 3.4 | JS |
| Sep 6  | 350.392 | 3.5 | RP |
| Sep 7  | 351.337 | 2.1 | JS |
| Sep 7  | 351.359 | 7.2 | RP |
| Sep 7  | 351.359 | 7.1 | IM |
| Sep 8  | 352.349 | 0.1 | JS |
| Sep 10 | 354.334 | 4.2 | JS |
| Sep 10 | 354.344 | 4.2 | DB |
| Sep 10 | 354.400 | 6.9 | IM |
| Sep 10 | 354.415 | 4.0 | RP |
| Sep 11 | 355.389 | 2.0 | RP |
| Sep 12 | 356.359 | 4.2 | DB |
| Sep 12 | 356.403 | 4.5 | RP |

**Table 4: Log of time-series observations**

| Date in 2007 (UT) | Time of maximum (JD-2454000) | Error (d) | Superhump cycle no. |
|---|---|---|---|
| Sept 2 | 345.5970 | 0.0006 | 0 |
| Sept 2 | 345.6896 | 0.0022 | 1 |
| Sept 2 | 345.7780 | 0.0013 | 2 |
| Sept 3 | 346.5845 | 0.0002 | 11 |
| Sept 3 | 346.6731 | 0.0002 | 12 |
| Sept 3 | 346.7620 | 0.0002 | 13 |
| Sept 3 | 346.8519 | 0.0004 | 14 |
| Sept 3 | 347.3840 | 0.0005 | 20 |
| Sept 3 | 347.3885 | 0.0002 | 20 |
| Sept 3 | 347.4760 | 0.0009 | 21 |
| Sept 3 | 347.4806 | 0.0010 | 21 |
| Sept 3 | 347.4764 | 0.0008 | 21 |
| Sept 4 | 348.3595 | 0.0002 | 31 |
| Sept 4 | 348.4502 | 0.0003 | 32 |
| Sept 7 | 351.3790 | 0.0002 | 65 |
| Sept 7 | 351.3825 | 0.0002 | 65 |
| Sept 7 | 351.4695 | 0.0003 | 66 |
| Sept 7 | 351.5577 | 0.0002 | 67 |
| Sept 7 | 351.6452 | 0.0002 | 68 |
| Sept 10 | 354.3929 | 0.0002 | 99 |
| Sept 10 | 354.4828 | 0.0004 | 100 |
| Sept 10 | 354.4804 | 0.0004 | 100 |
| Sept 10 | 354.5711 | 0.0002 | 101 |
| Sept 10 | 354.6601 | 0.0003 | 102 |

**Table 5: Times of superhump maximum**

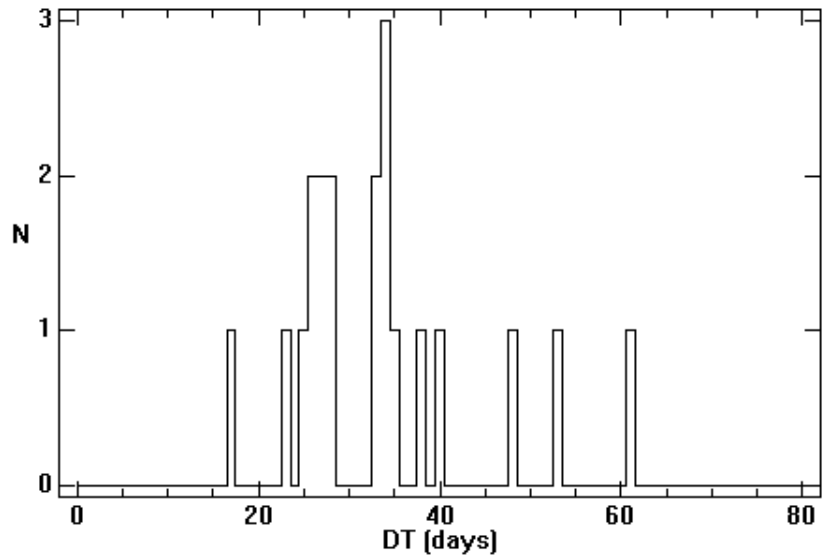

**Figure 1: Histogram of the outburst intervals in 1-day bins**

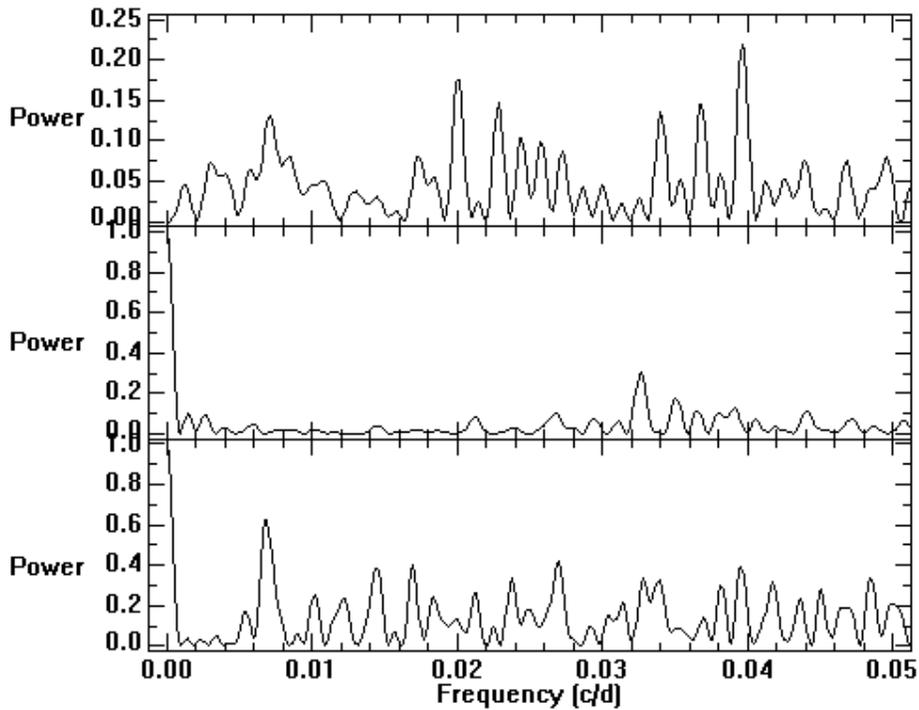

**Figure 2: (a-top) DFT power spectrum of the one-day means of the positive observations showing a series of features between f=0.02 and 0.04 c d$^{-1}$ due to the normal outbursts and another at f=0.007 c d$^{-1}$ due to the superoutbursts, (b-middle) window function of all the outburst timings showing a weak feature at f=0.033 c d$^{-1}$ and (b-bottom) the window function of the superoutburst timings showing a strong feature at f=0.007.**

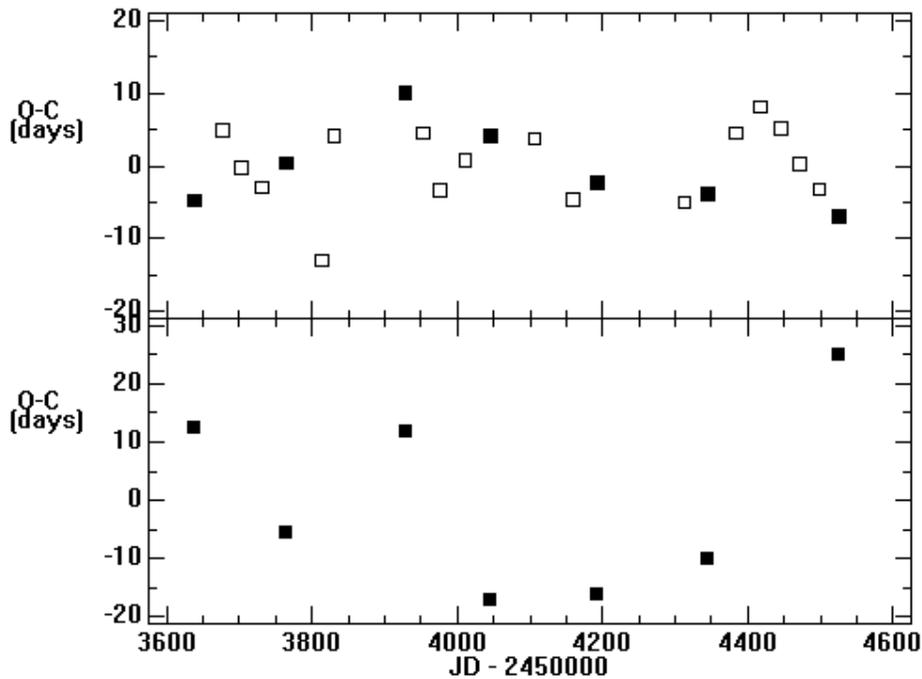

**Figure 3: O-C diagrams for (a-top) all the outbursts with the superoutbursts as filled symbols and (b-bottom) just the superoutbursts using equations (1) and (2) respectively.**

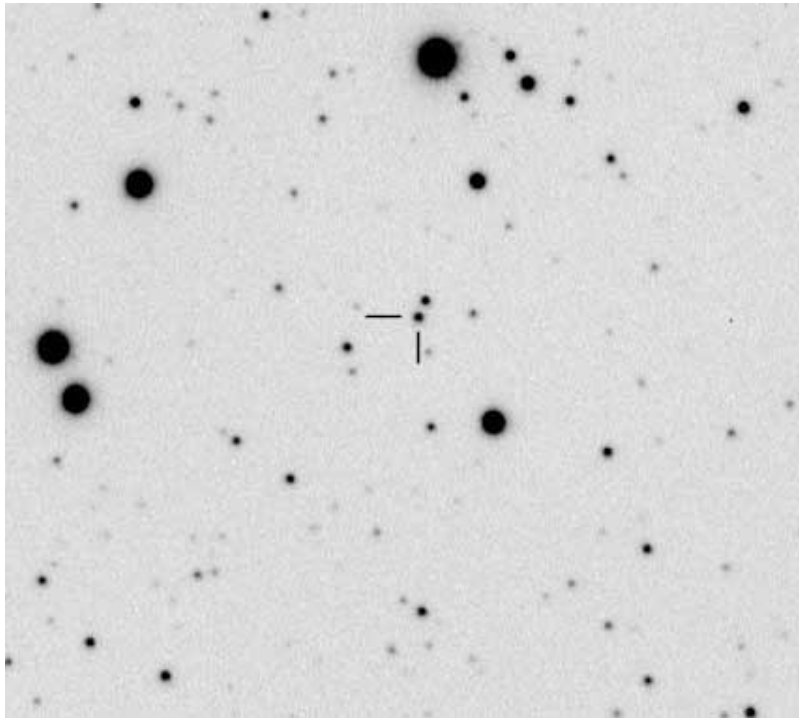

**Figure 4: V452 Cas in outburst at 15.7C on 2007 Sep 3.44**
7' x 7' with south at top

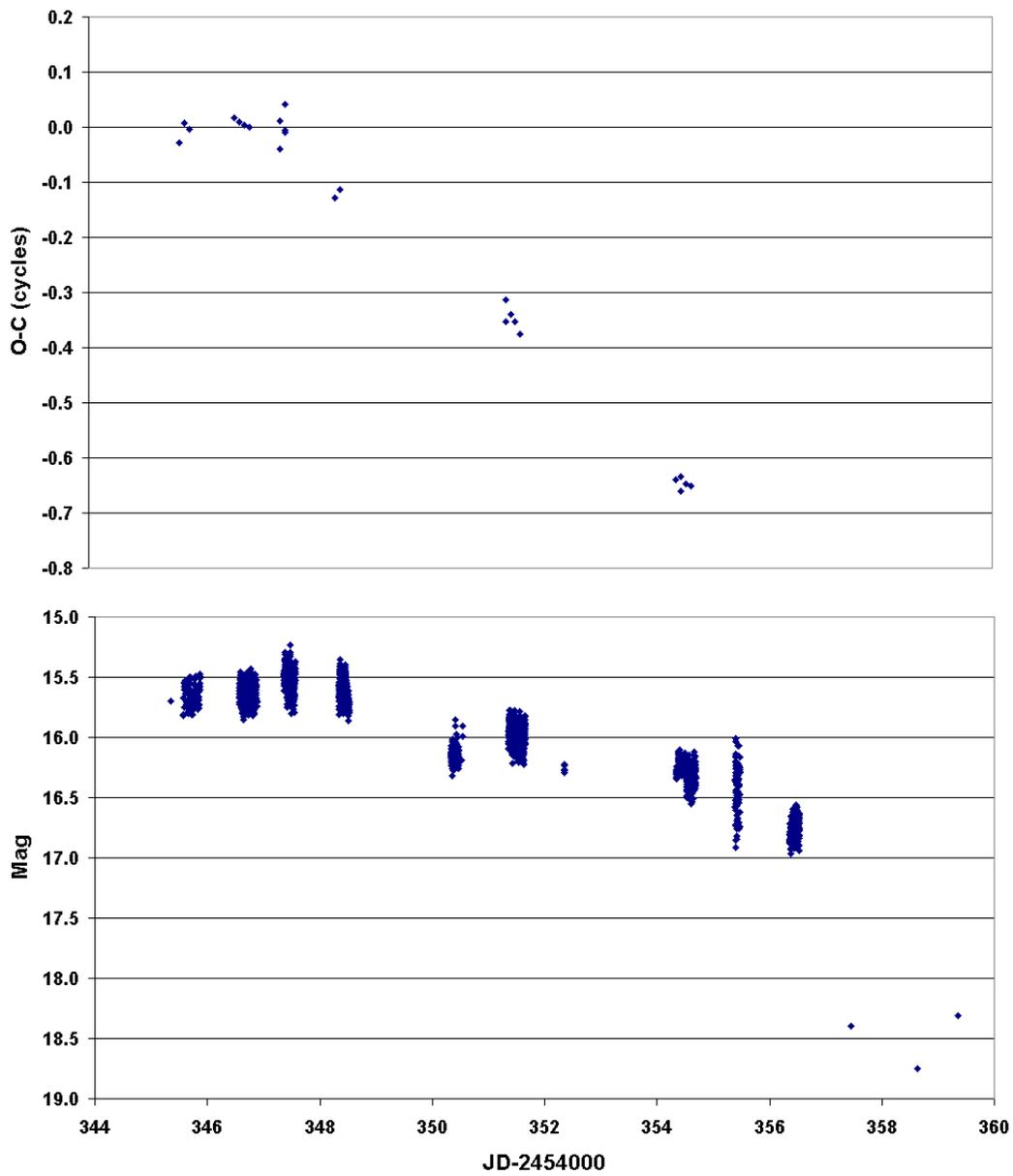

**Figure 5: Light curve of the 2007 September outburst (bottom) and O-C diagram of superhump maxima (top)**

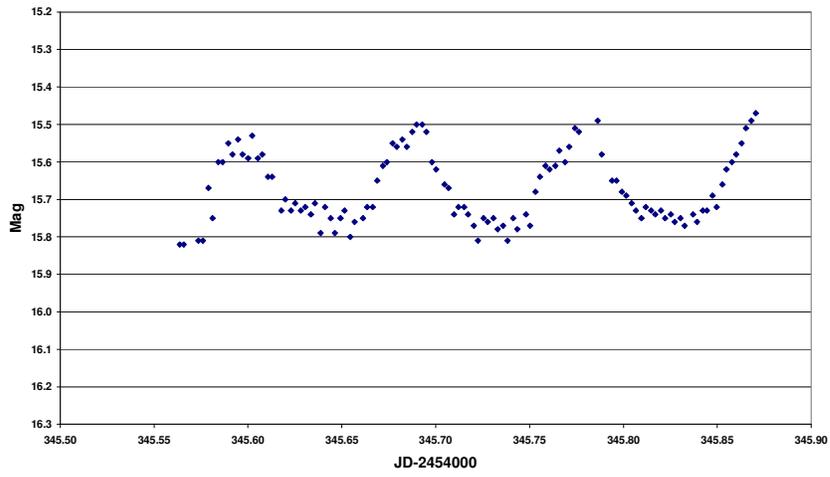

**Figure 6a: Time-series data from Sep 2 (JD 345)**

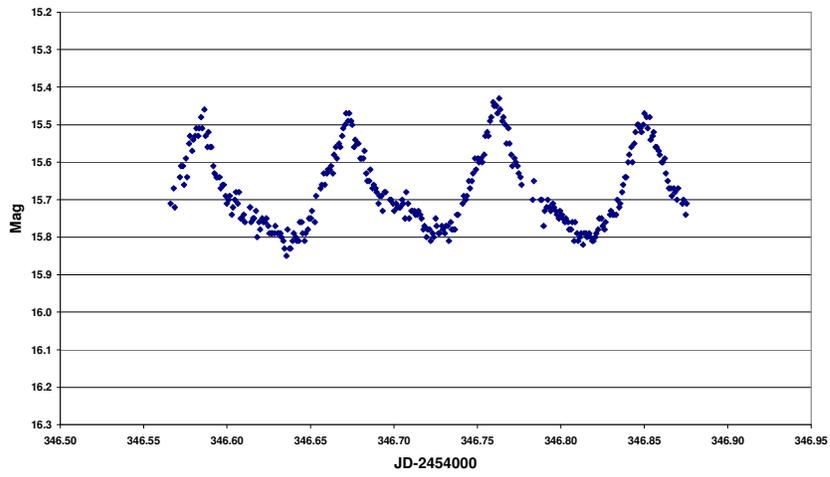

**Figure 6b: Time-series data from Sep 3 (JD 346)**

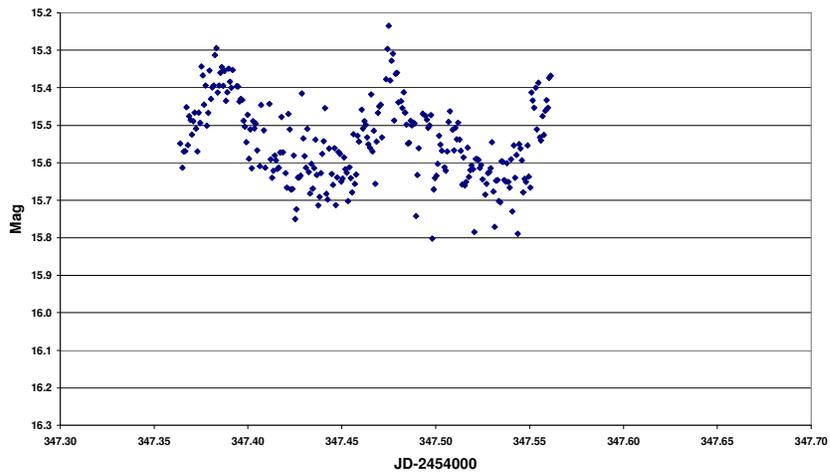

**Figure 6c: Time-series data from Sep 3 (JD 347)**

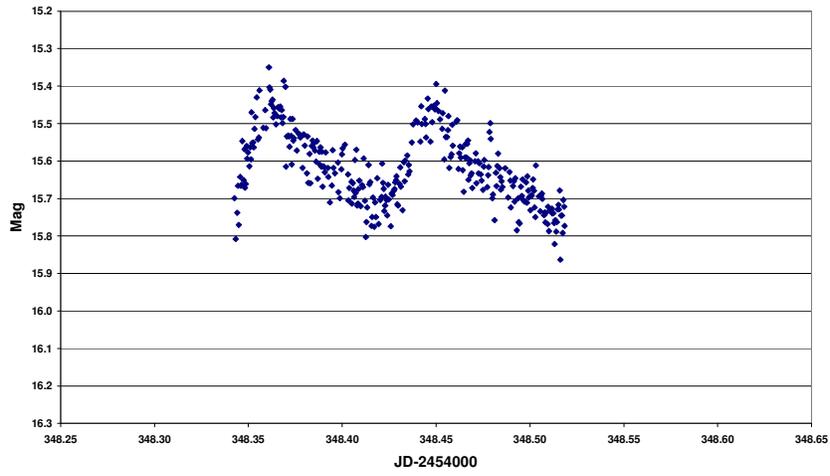

**Figure 6d: Time-series data from Sep 4 (JD 348)**

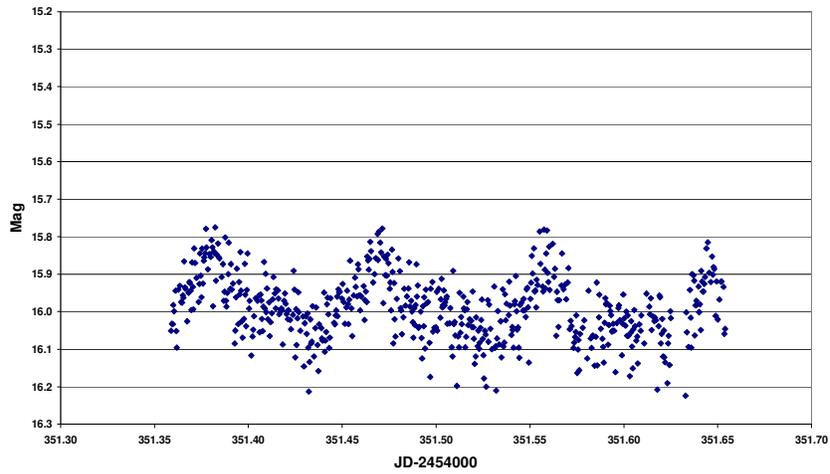

**Figure 6e: Time-series data from Sep 7 (JD 351)**

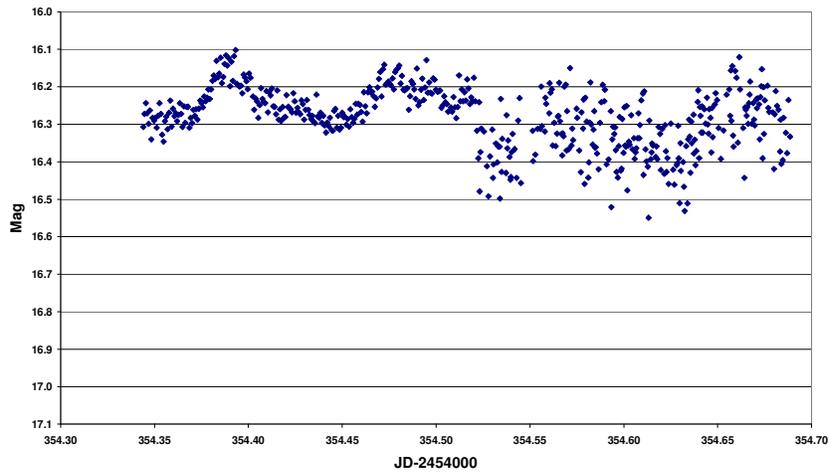

**Figure 6f: Time-series data from Sep 10 (JD 354)**
Note that the increased scatter in the right half of the plot is because these observations were made under poorer conditions

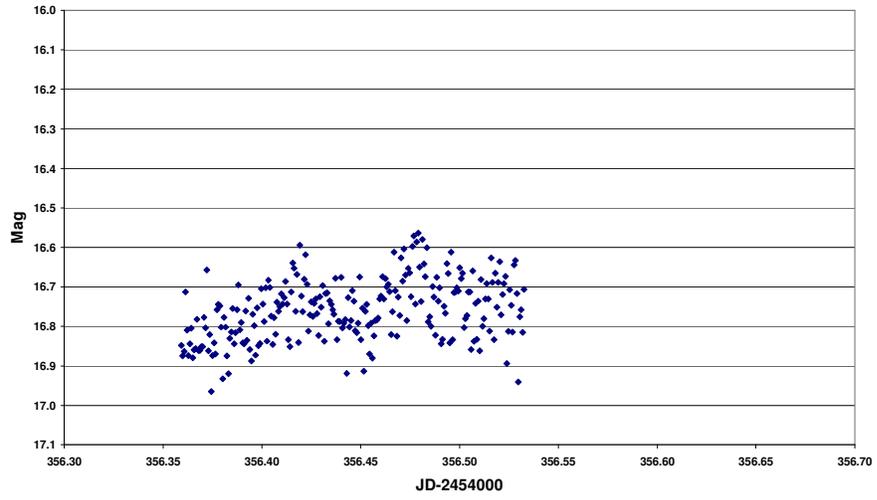

**Figure 6g: Time-series data from Sep 12 (JD 356)**

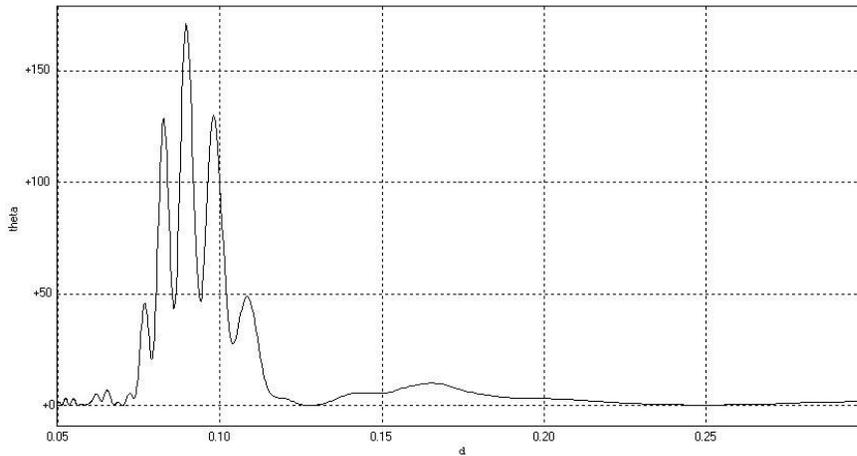
**Figure 7: Power spectrum of combined time-series data from JD 345 to 347**

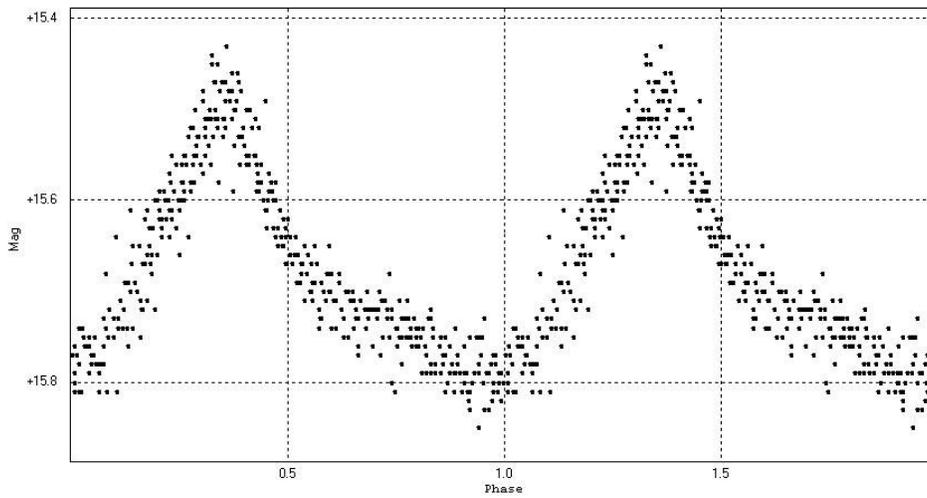
**Figure 8: Phase diagram of data from JD 345 to 347 folded on P$_{sh}$ of 0.0894 d**

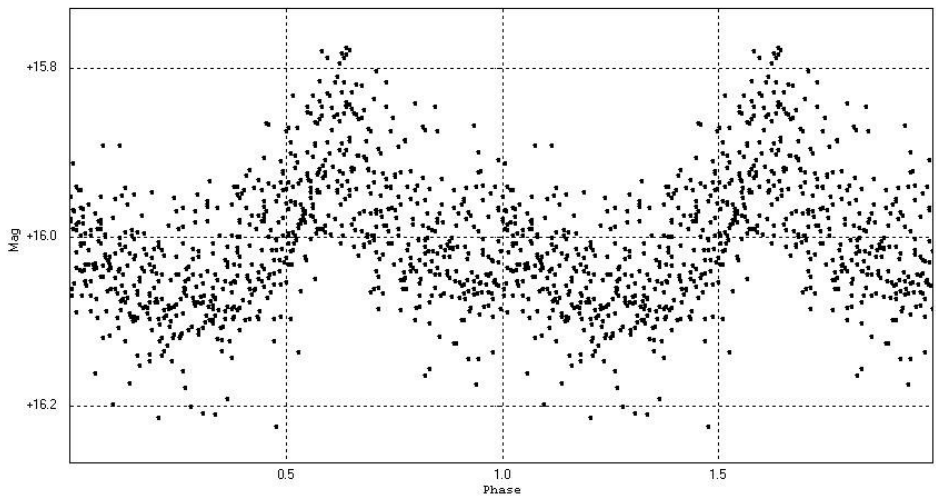
**Figure 9: Phase diagram of data from JD 348 to 354 folded on P$_{sh}$ of 0.0886 d**